\documentclass[12pt]{article}

\title{Cosmological constant and noncommutativity: A Newtonian point of view.\\}

\author{%
Juan M. Romero${}^{(1)}$\footnote{e-mail: sanpedro@nuclecu.unam.mx}
and J. A. Santiago${}^{(2)}$\footnote{e-mail: sgarciaj@uaeh.reduaeh.mx} 
\vspace{1cm}\\
${}^{(1)}${\it Instituto de Ciencias Nucleares}\\
{\it Universidad Nacional Aut\'onoma de M\'exico}\\
{\it Apartado Postal 70-543, M\'exico, DF, MEXICO}
\vspace{.5cm}\\
${}^{(2)}${\it Centro de Investigaci\'on Avanzada en Ingenier\'\i a Industrial}\\
{\it Universidad Aut\'onoma del Estado de Hidalgo}\\
{\it Apartado Postal 42-090, Pachuca, MEXICO}
}

\usepackage[dvips]{graphicx}

\usepackage[latin1]{inputenc}
\date{}
\begin{document}

\pagestyle{plain}

\maketitle

\begin{abstract}
We study a Newtonian cosmological model in the context of a noncommutative
space. It is shown that the trajectories of a test particle undergo 
modifications such that it no longer satisfies the cosmological
principle. For the case of a positive cosmological constant, 
spiral trajectories are obtained and corrections to the Hubble constant 
appear. It is also shown that, in the limit of a strong noncommutative
parameter, the model is closely related to a particle in a G\"odel-type 
metric.\\
\end{abstract}

\newpage

\section{Introduction}
\label{Intro}

It is well known that Newtonian gravity is no-relativistic, however, from
it one can get to conclusions which fully agree with general relativity.
For instance, J.~Michell (1784)~\cite{michell:gnus} and P.~S.~Laplace 
(1796)~\cite{laplace:gnus} have shown that the idea of black hole can be 
obtained from pure Newtonian gravity; though black holes do require
general relativity to be completely justified. On the other hand, in a 
cosmological level E. A. Milne (1934)~\cite{milne:gnus} did show that, 
from fluid equations and the cosmological principle, the Friedmann equation 
for the case of pressureless matter can be deduced. This approach is still
valid when pressure is much less than the energy density; otherwise general
relativity must be used \cite{zeldovich:gnus}.
Extensions of this approach can be found in Ref.~\cite{esos:gnus}. 
Moreover, the same equation can be obtained even from Newton's second law
~\cite{weinberg:gnus}. For a recent discussion of these topics see e.g. 
Refs.~\cite{myung:gnus,gibbons:gnus}. A great advantage of Newtonian gravity 
is its simplicity as it is just an 
external potential in a flat space. This simplicity allows one to study 
concepts otherwise difficult to understand from general relativity.\\

A difficult concept to introduce in general relativity is that of a
noncommutative (NC) space, however it can be incorporated in other 
field theories where it has proved useful for exploring possible 
modifications to the physics of small scales. The seminal idea of this 
proposal has been attributed to Heisenberg \cite{heisenberg:gnus}. NC 
spaces appear in a natural fashion in the context of string theory under 
some backgrounds \cite{witten:gnus}. Moreover, one can also construct in an 
independent manner a field theory in a NC space \cite{szabo:gnus}. The way 
to build this field theory starts by changing in the action the usual 
product of functions by a deformed product of the form
\begin{eqnarray}
(f\star g)(x)=f(x)g(x)+\theta \frac{i}{2}\{f(x),g(x)\}_{PS}
+{\cal O}(\theta^{2}),\label{eq:moyal}
\end{eqnarray}
where $\{.,.\}_{PS}$ denotes a Poisson structure and $\theta$
is a parameter controlling deformation such that when it is 
negligible the commutative limit is regained \cite{castellani:gnus}.
By considering the usual Poisson structure and making $\theta=\hbar$,
the usual field theories are obtained. However, if one considers 
alternative Poisson structures one gets to new theories. For instance, 
taking
\begin{eqnarray}
\{ x_{i}, x_{j}\}= \Theta_{ij}, \label{eq:poisson0}
\end{eqnarray}
with $\Theta_{ij}$ a constant, real and antisymmetric tensor; and
requesting associativity of the product, then the Moyal product
\begin{eqnarray}
f(x)\star g(x)= e^{i\hbar\Theta_{ij}\partial_{x_{i}}\partial_{y_{i}}} 
f(x)g(y)|_{x=y}, \label{eq:moyal2}
\end{eqnarray}
is obtained. In particular, the noncommutativity relation
\begin{eqnarray}
[x_{i},x_{j}]_{\star}=x_{i}\star x_{j}-x_{j}\star x_{i}=  i\hbar\Theta_{ij},
\end{eqnarray}
holds. An interesting property of field theories in these spaces is the
mixing of infrared and ultraviolet (UV/IR) divergences \cite{minwalla:gnus}.
This divergence mixing implies that the physics at large distances is not
disconnected from the physics at short scales. Several interactions have
been investigated in this context, for example, for the case of the
Standard Model see Ref. \cite{chaichian1:gnus}. Nevertheless, despite 
considerable progress, the task of accomplishing a final version of
NC general relativity has not been completed yet; though versions of
topological models with torsion already exist. For a detailed description
see Ref. \cite{compean:gnus} and references therein.\\

In this work we introduce noncommutativity in a cosmological Newtonian model
and study its consequences. To this end we first construct a NC
classical mechanics. Starting from the Poisson structure in Eq. 
(\ref{eq:poisson0}), we extend to the phase space as
\begin{eqnarray}
\{ x_{i}, x_{j}\}= \Theta_{ij},\quad \{ x_{i}, p_{j}\}=
 \delta_{ij},\quad \{ p_{i}, p_{j}\}= 0. \label{eq:poisson}
\end{eqnarray}
Notice that changing to a commutator this Poisson structure, the commutation
rules of a NC quantum mechanics are obtained \cite{chaichian:gnus}. However
by using Eq. (\ref{eq:poisson}) a NC classical mechanics can also be 
constructed. For example, with a Hamiltonian of the form
\begin{equation}
H=\frac{p_ip^i}{2m}+V(x),
\end{equation}
one gets the equations of motion
\begin{eqnarray}
&\dot x_{i}&=\frac{p_{i}}{m}+\Theta_{ij}\frac{\partial V}{\partial x^{j}},
\label{eq:,}\\
&\dot p_{i}&=-\frac{\partial V}{\partial x^{i}},
\label{eq:h1}
\end{eqnarray}
which yields,
\begin{equation} 
m\ddot x_{i}=-\frac{\partial V}{\partial x_{i}}+
m\Theta_{ij}\frac{\partial^{2 } V } { \partial x_{j}\partial x_{k}}
\dot x_{k}. \label{eq:..} 
\end{equation}
Notice that there is a correction to Newton's second law that depends
on the NC $\Theta$ parameter and on variations of the
external potential $V(x)$. This correction term can be regarded as 
a perturbation to the space due to the external potential. Eq. (\ref{eq:..})
has interesting properties itself. One of them is that the $t\to -t$ 
symmetry is broken unless the NC parameter is also 
transformed as $\Theta \to -\Theta$. A similar phenomenon occurs in
field theory \cite{jabbari1:gnus}. Also the rotational symmetry for
a central potential is broken. It can be shown that for the Kepler
potential Eq. (\ref{eq:..}) yields a perihelion shift of the planets.
In particular, for the case of Mercury this shift imposes a bound
to the noncommutativity scale of the order of $ 10^{15}GeV$ 
\cite{david1:gnus}. This is a remarkable bound as the lowest one
previously found was of the order of $ 10^{17}GeV$, given by 
$NC$-$QCD$ \cite{carone:gnus}. A detailed study of Eq. (\ref{eq:..})
can be found in Refs. \cite{david1:gnus,david:gnus} and some other
aspects of a NC classical mechanics in Ref. \cite{bola:gnus}.\\

The main aim of this letter is to study Eq. (\ref{eq:..})
for a Newtonian cosmological model. The first outcome is that 
a NC space cannot be consistent with the cosmological
principle as there is a privileged direction. Both, the cases
with positive and negative cosmological constant $\Lambda$ are 
discussed. For a negative $\Lambda$, the solutions obtained 
oscillate about the origin with a frequency depending on $\Theta$.
In the positive $\Lambda$, but otherwise commutative case, the 
trajectories are straight lines in which the particle departs
from the origin exponentially with time. However, in the 
NC case, the trajectories are spirals in which the
particle distance to the origin also grows exponentially with time 
but now $\Theta$ introduces small oscillations. The contribution
of these oscillations to the distance become important just after
every time period $T\propto 1/\Theta$; and in this period the distance
$r\propto e^{T}= e^{1/\Theta}$ is traveled. I.e. noncommutativity
at small distances ($\Theta$ small) produces effects at large 
distances, and therefore it connects short and large distances.
This phenomenon is analogous to the mixing of UV/IR 
divergences in field theory. Also, as it will be shown shortly,
the corrections to the distance $r$ give raise to corrections 
in the Hubble constant of the model. Finally, in the 
strong $\Theta$ limit, it is shown that the model is closely 
related to a particle in a G\"odel-type metric and that
the trajectories are circles which at the quantum mechanical
level have quantized radius.\\ 

This work is organized as follows. In section 2 the Newtonian
limit of general relativity without matter but with cosmological 
constant is briefly reviewed. In section 3 the NC case is
studied. The strong $\Theta$ limit is considered in section 4
and finally in section 5 the results are summarized.

\section{$AdS$ and $dS$ in the Newtonian limit} 
Einstein's equations with cosmological constant, $\Lambda$,
in vacuum are
\begin{eqnarray}
R_{\mu\nu}-\frac{1}{2}g_{\mu\nu}R=\Lambda g_{\mu\nu}.
\end{eqnarray}
Their solutions for maximum symmetry are of the form
\begin{eqnarray}
ds^{2}=-\left(1-\frac{\Lambda r^{2}}{3}\right)c^{2}dt^{2}
+\frac{dr^{2}}{\left(1-\frac{\Lambda r^{2}}{3}\right)}
+r^{2}d\Omega^{2}.\label{eq:veloz}
\end{eqnarray}
For negative $\Lambda$ this is the {\it anti de Sitter} space
and for positive $\Lambda$ is the {\it de Sitter} space. In
the Newtonian limit, $c\to\infty$ and  $\Lambda\to 0$ but 
$c\Lambda$ finite, the metric in Eq. (\ref{eq:veloz}) generates 
the potential
\begin{eqnarray}
V=-\frac{\Lambda mc^{2}}{6}r^{2}.  \label{eq:alo0}
\end{eqnarray}
The Hamiltonian of the system is
\begin{eqnarray}
H=\frac{p^{i}p_{i}}{2m}-\frac{\Lambda mc^{2}}{6}r^{2}.
\label{eq:hamiltoniano}
\end{eqnarray}
Assuming the usual Poisson structure the equations of motion
\begin{eqnarray}
m\frac{d^{2}\vec r }{dt^{2}} =m\frac{\Lambda c^{2}}{3}\vec r,
\label{eq:gib}
\end{eqnarray}
are obtained.
For negative $\Lambda$ there is an attractive force and the particle
oscillates about the origin with a distance (to the origin) given
by $r_{\Lambda <0}(t)=A|\sin{\sqrt{\Lambda}t}|$; but for positive
$\Lambda$ the force is repulsive and the particle departs exponentially
fast from the origin, i.e. $r_{\Lambda >0}(t)=Be^{\sqrt{\Lambda}t}$. 

Hubble's constant $H$ can be defined by the formula
\begin{eqnarray}
\dot r=Hr,
\end{eqnarray}
which for $\Lambda >0$ yields $H=\sqrt{\Lambda}$. For a study of
the symmetries of Eq. (\ref{eq:gib}) see Ref. \cite{gibbons:gnus}.\\

\section{Noncommutative case}
Let us now study the same system as in the previous section, but now
introducing the Poisson structure from Eq. (\ref{eq:poisson}).
Firstly, we shall see that a NC space is not compatible
with the cosmological principle. For that, as $\Theta_{ij}$ is
a $3\times 3$ antisymmetric matrix, it can be written as 
$\Theta_{ij}=\epsilon_{ijk}\Theta_{k}$. Therefore, $\vec\Theta$
defines a privileged direction.
For instance, if this vector is taken along the $z$ direction
(i.e. $\Theta_{k}=\delta_{k3}\Theta$) and Poisson brackets from
Eq. (\ref{eq:poisson}) are assumed, we obtain the relationships
\begin{eqnarray}
\{x_{1},x_{2}\}=\Theta, \quad \{x_{2},x_{1}\}=-\Theta, 
\quad \{x_{3},x_{i}\}=0.
\end{eqnarray}
That is, noncommutativity only affects the plane perpendicular
to the $z$ direction. This defines a privileged direction
and in such a space the cosmological principle is not possible.\\

Let us now turn to the Hamiltonian in Eq. (\ref{eq:hamiltoniano})
with the Poisson structure from Eq. (\ref{eq:poisson}). By taking
$\Theta_{ij}=\epsilon_{ij3}\Theta$ one gets to the equations of
motion 
\begin{eqnarray}
&m\ddot x&=\frac{m\Lambda c^{2}}{3}x-\frac{\Lambda c^{2}}{3}
m^{2}\Theta \dot y,\label{eq:primera}\\
&m\ddot y&=\frac{m\Lambda c^{2}}{3}y+\frac{\Lambda c^{2}}
{3}m^{2}\Theta \dot x,\label{eq:segunda}\\
&m\ddot z&=\frac{m\Lambda c^{2}}{3}z.\label{eq:tercera}
\end{eqnarray}
These equations are essentially identical to
those for a harmonic oscillator in a 
constant magnetic filed along the $z$ direction. As expected,
movement along the $z$ direction is not affected by noncommutativity,
but it is in the plane perpendicular to $z$. This is the reason
behind being unable to take solutions of the form 
$\vec r=R(t)\vec r(t_{0})$, which are consistent with the
cosmological principle.\\

\begin{figure}[h]
\begin{center}
\framebox{
\includegraphics[width=2in,height=2in,angle=0]{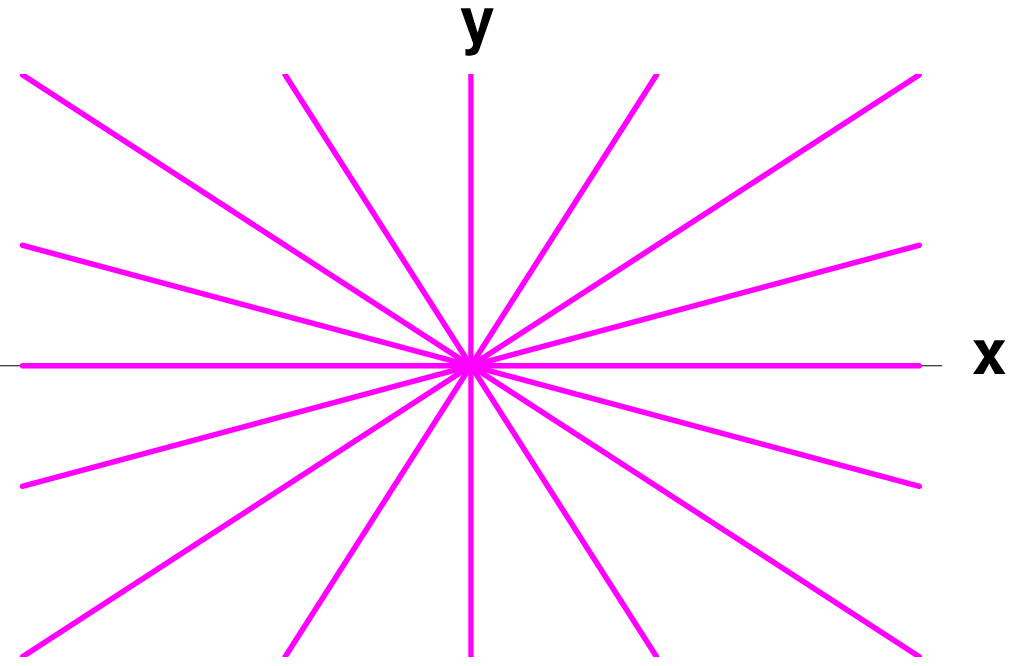}}
\framebox{
\includegraphics[width=2in,height=2in,angle=0]{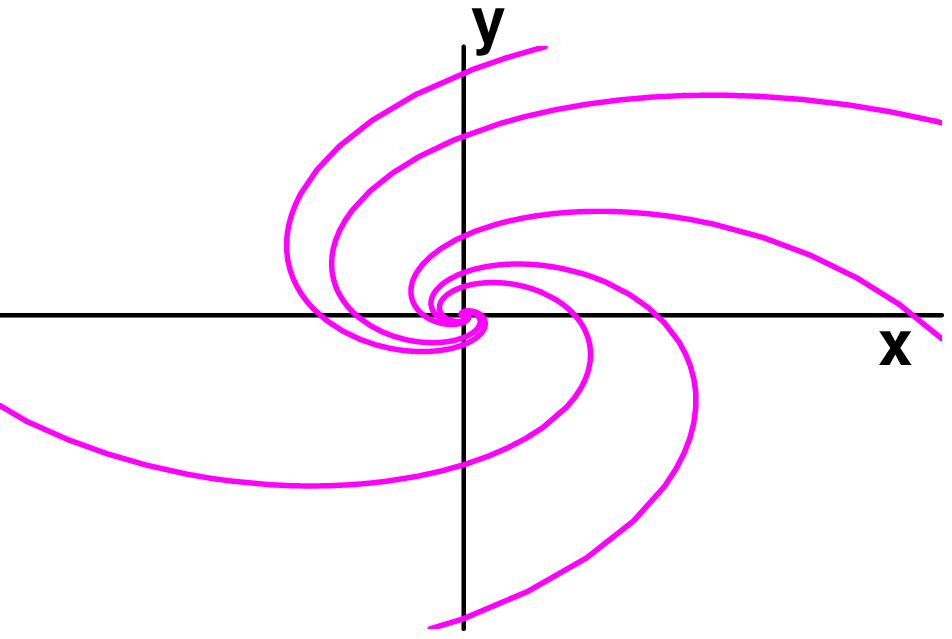}}
\end{center}
\caption{\small Trajectories of a particle in a space with positive
cosmological constant. Left figure represents the commutative case. 
Right figure is for the noncommutative one.}\label{efe}
\end{figure}

As the $\Theta$ parameter affects only the $x$-$y$ plane, for
simplicity (unless stated otherwise), we will assume the solution
$z(t)=0$ along the $z$ direction. Now, looking for solutions of
the form $x=x_{0}e^{\omega t}$ and $y=y_{0}e^{\omega t}$, we find
that
\begin{eqnarray}
&\omega&=\sqrt{\alpha}\left[\left(1-\frac{\beta^{2}}{2\alpha}\right)
\pm \sqrt{\left(1-\frac{\beta^{2}}{2\alpha}\right)^{2}-1}\,\,\right]^{1/2},\\
&y_{0}&=x_{0}\frac {\beta \omega}{\alpha-\omega^{2}},
\end{eqnarray}
with $\alpha=\Lambda c^{2}/3$ and $\beta=m\Theta\alpha$. To second order
in $\Theta$ this yields
\begin{eqnarray}
\omega=\sqrt{\alpha}\left( 1-\frac{\beta^{2}}{4\alpha}
\mp \frac{\beta}{2\alpha}\sqrt{-\alpha}\right).\label{eq:frecuencia}
\end{eqnarray}
It can be seen from Eq. (\ref{eq:frecuencia}) that if $\Lambda <0$
the movement is completely oscillatory as in the commutative case.
However, the oscillations now get corrected by the NC
parameter. That is, the cosmological constant gets a correction from
$\Theta$. On the other hand, if $\Lambda >0$ the frequencies are no 
longer real as they get an imaginary correction from $\Theta$ of
the form
\begin{eqnarray}
\omega=\omega_{R}\pm \omega_{I}=\sqrt{\alpha}
\left( 1-\frac{\beta^{2}}{4\alpha}\right) \pm i\frac{\beta}{2},
\end{eqnarray}
so that a test particle now departs from the origin and it also
oscillates in the $x$-$y$ plane, in contrast to the commutative 
case. To this order, linearly independent trajectories in the 
$x$-$y$ plane are
\begin{eqnarray}
&(x,y)&=e^{\omega_{R}t}\left(\cos(\omega_{I}t),\left(1-\frac{\beta^{2}}
{8\alpha}\right)\sin(\omega_{I}t)\right),\label{eq:espiral1}\\
&(x,y)&=e^{\omega_{R}t}\left(\sin(\omega_{I}t),-\left(1-\frac{\beta^{2}}
{8\alpha}\right)\cos(\omega_{I}t)\right).\label{eq:espiral2}
\end{eqnarray}
Therefore, in this case the trajectories are no straight lines, but
spirals. This trajectories are schematically shown in Figure \ref{efe}.
From these solutions, Eqs. (\ref{eq:espiral1}) and (\ref{eq:espiral2}),
it can be seen that the cosmological principle is broken in the
NC plane.\\

\begin{figure}
\begin{center}
\includegraphics[width=6.0cm,height=6.0cm]{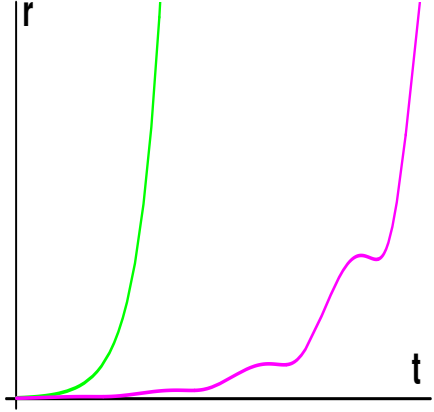} 
\end{center}
\caption{\textrm{{\protect\small Distance to the origin for the 
particle position as a function of time. Order zero and first 
order are represented by the line on the left. Second order
by the line on the right.}}}\label{esta3}
\end{figure}

On the other hand the particle distance to the origin at time $t$
is
\begin{eqnarray}
r(t)=Ae^{\sqrt{\omega_{R}}t}\sqrt{1-\frac{\beta^{2}}{4\alpha}
\cos^{2}\omega_{I}t}. 
\label{eq:animo}
\end{eqnarray}
To first order in $\Theta$ this grows exponentially. However, to second
order small oscillations appear. The behaviour of $r(t)$ is shown
schematically in Figure \ref{esta3}. The oscillations in $r(t)$ are 
analogous to the scale factor oscillations of a cosmological model
of a quasi-steady-state type \cite{hoyle:gnus,hoyle2:gnus}. \\

From Eq. (\ref{eq:animo}) it can be seen that the effect of the
noncommutativity in $r(t)$ is essentially suppressed, showing up
only when $\omega_{I}t=n\pi$. That is, when
\begin{eqnarray}
t=\frac{n\pi}{\omega_{I}}=\frac{3n\pi}{m\Theta \Lambda c^{2}}.
\label{eq:tiempo}
\end{eqnarray}
The distance traveled by the test particle during this time
is 
$$r(t=n\pi/\omega_{I} )\propto e^{\left(\frac{3\pi n}{m\sqrt{\Lambda 
c^{2}}}\right)\frac{1}{\Theta} }.$$
 Now, taking the limit $\Theta\to 0$,
then $r\to\infty$. Thus, for small $\Theta$ the first oscillation
will appear at a very long distance and this can be interpreted 
as that noncommutativity at short distances has effects at very
long ones; i.e. short and very long distances are no disconnected.
This phenomenon is analogous to the mixing of the UV/IR
divergences appearing in a field theory in a NC
space. Notice that for this phenomenon to be observed, corrections
to second order in $\Theta$ have to be considered. Also note that
the time defined in Eq. (\ref{eq:tiempo}) is proportional to
the time the particle requires to turn $n$ times about the origin.

Now, by defining Hubble's constant from $\dot r=H r$, for the 
$\Lambda >0$ case we find
\begin{eqnarray}
H=\sqrt{\omega_{R}}=\sqrt{\alpha}
\left(1-\frac{\beta^{2}}{4\alpha}\right).
\end{eqnarray}  
Therefore, to this order a constant correction to $H$ appears. 
To higher orders Hubble's constant gets corrections that depend 
on time.\\

\section{Strong $\Theta$ limit}

The strong $\Theta$ limit in Eqs. (\ref{eq:primera})-(\ref{eq:tercera})
can be obtained as $\left(m\Lambda c^{2}/3\right)\to 0$ and 
$\Theta \to \infty$, but $\left(m\Lambda c^{2}/3\right)m\Theta\,$
finite, so that
\begin{eqnarray}
&m\ddot x&=-\frac{\Lambda c^{2}}{3}m^{2}\Theta \dot y,\label{eq:esas1}\\
&m\ddot y&=\frac{\Lambda c^{2}}{3}m^{2}\Theta \dot x, \\
&m\ddot z&=0.\label{eq:esas2}
\end{eqnarray}
Eqs. (\ref{eq:esas1})-(\ref{eq:esas2}) can also be regarded as the
equations of motion of a charged particle in a magnetic field along the 
$z$ direction. In such a case, $\Lambda$ plays the role of the charge
and $\Theta$ the magnetic field. If the particle speed along the $z$
direction is zero, then it describes a circle centred in $(x_{0},y_{0})$. 
Whether the movement is clock or anticlockwise depends on the sign of 
$\Lambda$. Moreover, as in the quantum formalism the commutation 
relations,
\begin{eqnarray}
[\hat x, \hat y]=i\hbar \Theta, 
\end{eqnarray}
hold and therefore the radius $\hat r$ of the circles is quantized,
\begin{eqnarray}
\hat r^{2}=2\hbar \Theta \left(n+\frac{1}{2}\,\right);
\end{eqnarray}
just as it happens in quantum mechanics to a charged particle in 
a uniform magnetic field \cite{laquefalta:gnus}.\\

As explained in Sec. \ref{Intro}
the commutative version of this model yields results in agreement
with general relativity. Assuming that analogously happens for
the noncommutative case, we could say that if at the beginning of 
the universe matter was dominated by a cosmological constant 
(i.e. vacuum) and the space is NC with $\Theta$ strong, then at 
the quantum level a test particle would describe helices of 
quantized radius; but as $\Theta$ becomes weak, the circles 
get deformed.\\

Now, by considering the metric
\begin{eqnarray}
ds^{2}=-\left(dt+\frac{{\bf \Omega}\sinh^{2} l\rho}{l^{2}} 
d\phi \right)^{2}+\frac{\sinh^{2} 2l\rho}{4l^{2}} d\phi^{2}
+d\rho^{2}+dz^{2}, \label{eq:metri}
\end{eqnarray}
taking the limit $l\to 0$, and expressing it in terms of the
coordinates
\begin{equation}
x=\rho \cos\phi, \quad y=\rho \sin\phi,
\end{equation}
then the geodesics of this metric are of the same form as the
equations of motion (\ref{eq:esas1})-(\ref{eq:esas2}); where
$\bf{\Omega}$ plays the role of the $\Theta\left(\Lambda c^{2}/3\right)$
parameter. Since for the $l^{2}=2{\bf \Omega}$ case one has the G\"odel
metric \cite{godel:gnus}, the metric in Eq. (\ref{eq:metri}) is called
a G\"odel-type metric and is the solution to Einstein's equations
with cosmological constant and nonzero energy momentum tensor. 
For more properties of this metric see e.g. Ref. \cite{reboucas:gnus}. 
An interesting relationship between this space and Landau's problem can 
be found in Ref. \cite{simon:gnus}.\\

Another way of getting noncommutativity is through a Matrix model.
In this formalism it is also possible to propose a Newtonian
cosmological model \cite{alvarez:gnus}. Its worth mentioning
that recently new observations have been obtained suggesting
the existence of a positive cosmological constant 
\cite{peebles:gnus}.

\section{Summary}
In this letter a Newtonian cosmological model in a NC 
classical mechanics is studied. As in a NC space there are 
privileged directions, the trajectories of a test particle 
are not compatible with the cosmological principle. If the 
cosmological constant is negative, the trajectories are 
oscillatory, with oscillation frequency depending on the NC 
parameter. On the contrary, if the cosmological constant is 
positive, the trajectories on the NC plane are spirals and to
first order in $\Theta$ the distance from the origin to the
particle, $r(t)$, grows exponentially fast. To this order 
Hubble's constant is the same as for the commutative case. 
To second order, however, $r(t)$ also grows exponentially 
but develops small oscillations. In this approximation Hubble's
constant gets corrections. The oscillation period in $r(t)$,
$T\propto 1/\Theta$, so for $\Theta$ small this period is very
long. On the other hand, because $r(t)$ grows exponentially
with time, the distances at which perturbations can be observed
are very long. Therefore, considering noncommutativity at
short distances has implications at large distances. Finally, 
it is shown that in the strong $\Theta$ limit the trajectories 
are circles, whose radius are discretized at the quantum level. 
In this limit there is a relation with a G\"odel-type metric.

\section{Acknowledgments}
We would like to thank J.~David Vergara for useful suggestions and 
Adolfo Zamora for critical reading and useful comments to improve
the manuscript. JAS acknowledges financial support from the
PROMEP program under grant PTC-260.

\end{document}